\documentclass[onecolumn,showpacs,showkeys,preprintnumbers]{revtex4}
\usepackage{amssymb}

\usepackage{amsmath}
\usepackage{graphicx}

\begin{document}
\title{Current Observational Constraints to Holographic Dark Energy Model with New Infrared cut-off
 via Markov Chain Monte Carlo Method}
\author{Yuting Wang}
\author{Lixin Xu}
\email{lxxu@dlut.edu.cn}
\affiliation{School of Physics and Optoelectronic Technology,\\
Dalian University of Technology, Dalian, Liaoning 116024, P. R.
China}
\begin{abstract}
In this paper, the holographic dark energy model with new infrared
(IR) cut-off for both the flat case and the non-flat case are
confronted with the combined constraints of current cosmological
observations: type Ia Supernovae, Baryon Acoustic Oscillations,
current Cosmic Microwave Background, and the observational hubble
data. By utilizing the Markov Chain Monte Carlo (MCMC) method, we
obtain the best fit values of the parameters with $1\sigma, 2\sigma$
errors in the flat model: $\Omega_{b}h^2=0.0233^{+0.0009
+0.0013}_{-0.0009 -0.0014}$, $\alpha=0.8502^{+0.0984
+0.1299}_{-0.0875 -0.1064}$, $\beta=0.4817^{+0.0842
+0.1176}_{-0.0773 -0.0955}$, $\Omega_{de0}=0.7287^{+0.0296
+0.0432}_{-0.0294 -0.0429}$, $\Omega_{m0}=0.2713^{+0.0294
+0.0429}_{-0.0296 -0.0432}$, $H_0=66.35^{+2.38 +3.35}_{-2.14
-3.07}$. In the non-flat model, the constraint results are found in
$1\sigma, 2\sigma$ regions: $\Omega_{b}h^2=0.0228^{+0.0010
+0.0014}_{-0.0010 -0.0014}$, $\Omega_k=0.0305^{+0.0092
+0.0140}_{-0.0134 -0.0176}$, $\alpha=0.8824^{+0.2180
+0.2213}_{-0.1163 -0.1378}$, $\beta=0.5016^{+0.0973
+0.1247}_{-0.0871 -0.1102}$, $\Omega_{de0}=0.6934^{+0.0364
+0.0495}_{-0.0304 -0.0413}$, $\Omega_{m0}=0.2762^{+0.0278
+0.0402}_{-0.0320 -0.0412}$, $H_0=70.20^{+3.03 +3.58}_{-3.17
-4.00}$. In the best fit holographic dark energy models, the
equation of state of dark energy and the deceleration parameter at
present are characterized by $w_{de0}=-1.1414\pm0.0608,
q_0=-0.7476\pm0.0466$ (flat case) and $w_{de0}=-1.0653\pm0.0661,
q_0=-0.6231\pm0.0569$ (non-flat case). Compared to the $\Lambda
\textmd{CDM}$ model, it is found the current combined datasets do
not favor the holographic dark energy model over the $\Lambda
\textmd{CDM}$ model.
\end{abstract}

\maketitle

\section{Introduction}
Since 1998, the type Ia supernova (SNe Ia) observations
\cite{ref:Riess98,ref:Perlmuter99} have shown that our universe has
entered into a phase of accelerating expansion. During these years
from that time, many additional observational results, including
current Cosmic Microwave Background (CMB) anisotropy measurement
from Wilkinson Microwave Anisotropy Probe
(WMAP)\cite{ref:Spergel03,ref:Spergel06}, and the data of the Large
Scale Structure (LSS) from Sloan Digital Sky Survey (SDSS)
\cite{ref:Tegmark1,ref:Tegmark2}, also strongly support this
suggestion. These observational results have greatly inspirited
theorists to understand the mechanism of the accelerating expansion
of the universe, which is usually attributed to an exotic energy
component with negative pressure, dubbed dark energy (DE). The
simplest but most natural candidate of DE is the cosmological
constant $\Lambda$, with the constant equation of state (EOS)
$w=-1$. As we know, the cosmic concordance model confronts with two
difficulties: the fine-tuning problem and the cosmic coincidence
problem. Both of these problems are related to the DE density. In
order to solve or alleviate cosmological constant puzzles, many
dynamical DE models are proposed, where the DE density and its EOS
are time-varying. However, the predictions of the cosmological
constant model still fit to the current observations
\cite{ref:LCDM1,ref:LCDM2,ref:LCDM3}. Therefore the dynamical DE
models being proposed should not be far away from the cosmological
constant model, such as quintessence
\cite{ref:quintessence01,ref:quintessence02,ref:quintessence1,ref:quintessence2,ref:quintessence3,ref:quintessence4},
phantom \cite{ref:phantom}, quintom \cite{ref:quintom}, K-essence
\cite{ref:kessence}, tachyon \cite{ref:tachyon}, ghost condensate
\cite{ref:ggc}, holographic DE \cite{ref:holo1,ref:holo2} and
agegraphic DE \cite{ref:age1,ref:age2} etc. Although many DE models
have been presented, the nature of DE is still a conundrum. This
puzzle can not be understood before a complete theory of quantum
gravity is established. But the two additional aspects from the
current cosmological observations and some basic quantum
gravitational principles may shed light on probing the nature of DE.

On the one hand, provided that we know little on the theoretical
nature of DE at present, the combined cosmic observations can play
an important role in understanding the nature of DE. The
cosmological parameters space in the DE model can be determined by
the constraints of the data combinations. Recently, the 397 SN Ia
data was compiled in Ref. \cite{ref:Condata} by adding CfA3 sample
from the CfA SN Group to the Union set by Ref. \cite{ref:Kowalski},
which include 250 SN Ia at high redshift but only 57 at low
redshift, to form the Constitution set. Aside from the SN Ia data,
the combined analysis is required in order to break the degeneracy
between the cosmological parameters, which includes cosmic
observations from baryon acoustic oscillations (BAO), CMB and the
observational Hubble data (OHD). The BAO are detected in the
clustering of the combined 2dFGRS and SDSS main galaxy samples or
the SDSS luminous red galaxies and measure the distance-redshift
relation. From these samples, the values of $[r_s(z_d)/D_V(0.2),
r_s(z_d)/D_V(0.35)]$ and their inverse covariance matrix in the
measurement of BAO can be obtained \cite{ref:Percival2}. For the
measurement of CMB, we utilize the shift parameter $R$ at the photon
decoupling epoch $z_\ast$, the acoustic scale $l_A(z_\ast)$, and
together with the physical baryon density parameter multiplied by
100, thus it is $100\Omega_bh^2$ \cite{ref:Komatsu2008, ref:Bueno
Sanchez}. Here, it is worth noting that the WMAP distance
information $R(z_\ast)$ and $l_A(z_\ast)$ can not be measured by
WMAP directly, but are derived from making a global fitting
constraint with MCMC method by using the full WMAP data on the
assumption that a certain cosmological model has been given in
advance \cite{ref:lAR}. Although in theory the inverse covariance
matrix on $R(z_\ast)$ and $l_A(z_\ast)$ is model dependent, it is
feasible to use the derived results about $R(z_\ast)$ and
$l_A(z_\ast)$ to constrain the parameters in another DE model since
$R(z_\ast)$ and $l_A(z_\ast)$ do not depend strongly on the DE model
which is not far away from the cosmological constant model
\cite{ref:lAR}. What is more, the paper \cite{ref:ywang} has been
demonstrated that $[R(z_\ast), l_A(z_\ast), 100\Omega_bh^2]$
effectively provide a good summary of CMB data when the DE model
parameters are constrained. In addition, we employ the OHD at twelve
different redshifts determined by using the differential ages of
passively evolving galaxies in Ref. \cite{ref:0907}, where the value
of the Hubble constant is replaced by $H_0=74.2\pm3.6$ in Ref.
\cite{ref:0905}, and add the three more observational data
$H(z=0.24)=79.69\pm2.32, H(z=0.34)=83.8\pm2.96,$ and
$H(z=0.43)=86.45\pm3.27$ in \cite{ref:0807}. Since the constraint
results of a given model are dependent on the combined data
\cite{ref:Gdata, ref:XUdata}, in this paper we use a fully combined
observations from the 397 SN Ia standard candle data, the value of
$[r_s(z_d)/D_V(0.2), r_s(z_d)/D_V(0.35)]$ and their inverse
covariance matrix in the measurement of BAO, the values of
$[R(z_\ast), l_A(z_\ast), 100\Omega_bh^2]$ and their inverse
covariance matrix in the measurement of CMB, and the fifteen OHD.

On the other hand, the models which are constructed in light of some
fundamental principle are more charming, since this kind of DE model
may exhibit some underlying features of DE, for instance the
holographic DE model \cite{ref:holo1,ref:holo2} and the agegraphic
DE model \cite{ref:age1,ref:age2}. The holographic DE model is built
on the basis of holographic principle and some features of quantum
gravity theory. The agegraphic DE model is derived from taking the
combination between the uncertainty relation in quantum mechanics
and general relativity into account. In this paper, we focus on the
holographic DE model, which is considered as a dynamic vacuum
energy. According to the holographic principle, the number of
degrees of freedom in a bounded system should be finite and is
related to the area of its boundary. By applying the principle to
cosmology, one can obtain the upper bound of the entropy contained
in the universe. For a system with size $L$ and UV cut-off $\Lambda$
without decaying into a black hole, it is required that the total
energy in a region of size $L$ should not exceed the mass of a black
hole of the same size, thus $L^3\rho_\Lambda\leq LM_{pl}^2$. The
largest $L$ allowed is the one saturating this inequality, thus we
obtain the holographic DE density
\begin{eqnarray}
&&\rho_\Lambda=\frac{3c^2M_{pl}^2}{L^2},
\end{eqnarray}
where c is a numerical constant and $M_{pl}$ is the reduced Planck
Mass $M_{pl}\equiv1/\sqrt{8\pi G}$. It just means a duality between
UV cut-off and IR cut-off. The UV cut-off is related to the vacuum
energy, and IR cut-off is related to the large scale of the
universe, for example Hubble horizon, particle horizon, event
horizon, Ricci scalar or the generalized functions of dimensionless
variables as discussed by
\cite{ref:holo1,ref:holo2,ref:holo3,ref:EPJCXU}. Next, we give a
brief review on the main results when Hubble horizon, particle
horizon, event horizon or Ricci scalar are taken as the IR cut-off,
respectively.

$\bullet$ $L^{-2}=H^2$. As pointed in \cite{ref:holo2}, it is found
that the holographic DE density is in proportion to $H^2$, the same
as dark matter density, i.e. $\rho_{de}/c^2=\rho_m/(1-c^2)\propto
H^2$. It appears that it is natural to solve the coincidence
problem. However, Hsu \cite{ref:holo1} pointed out that the dark
energy EOS $w_{de}=0$ was obtained in this instance. It is obvious
that this result is not consistent with the current observations.
This bad situation can be changed by considering the holographic DE
with Hubble horizon as the time variable cosmological constant. More
detailed analysis is presented in Ref. \cite{ref:XU071709}.

$\bullet$
$L^{-2}=R_{ph}(a)=a\int_0^t\frac{dt'}{a(t')}=a\int_0^a\frac{da'}{Ha'^2}$.
As shown in paper \cite{ref:holo2}, Li pointed out that this yields
the dark energy EOS is not less than $-1/3$. Thus the current
accelerated expansion of our universe can not be well explained.
However, this result in \cite{ref:holo2} is obtained on the
assumption that DE dominates. The holographic DE model with particle
horizon has been discussed in detail by \cite{ref:XU054772}.

$\bullet$
$L^{-2}=R_{eh}(a)=a\int_t^\infty\frac{dt'}{a(t')}=a\int_a^\infty\frac{da'}{Ha'^2}$.
The holographic DE model with event horizon can reveal the dynamic
nature of the vacuum energy and provide a desired EOS of the
holographic DE with the model parameter $c$. Furthermore, the
holographic DE behaves like quintessence, cosmological constant and
phantom respectively for the different values of the model
parameter: $c\geq1$, $c=1$ and $c\leq1$ \cite{ref:holo2}. Therefore,
the value of model parameter $c$ plays a crucial role in determining
the property of holographic DE in this case. However, this model is
confronted with the causality problem: why should the present
density of DE be determined by the future event horizon of the
universe.

$\bullet$ $L^{-2}=R=-6(\dot{H}+2H^2+\frac{k}{a^2})$. In
\cite{ref:holo3}, it has shown that this model can avoid the
causality problem and naturally solve the coincidence problem of
dark energy after Ricci scalar is taken as the IR cut-off and the
parameters have been well constrained by the combined astronomical
observations \cite{ref:MPLAXU,ref:LiRicci}.

Subsequently, In \cite{ref:holo4}, Granda and Oliveros generalized
the form of the IR cut-off on the basis of the Ricci scalar:
\begin{eqnarray}
&&L^{-2}=\alpha H^2+\beta \dot{H},
\end{eqnarray}
where there are two independent model parameters $\alpha$ and
$\beta$, which can be determined by using the combined constraints
of the thorough observational datasets. In this paper, we consider
the holographic DE model with new IR cut-off in both flat and
non-flat case. The performance of a global fitting will be made by
using the Markov Chain Monte Carlo (MCMC) method. In this way, we
can work in the framework of multi-parameter freedoms, including the
basic cosmological parameters ($\Omega_bh^2, \Omega_ch^2, \Omega_k$)
and the new-added model parameters ($\alpha, \beta$).

The paper is organized as follows. In next section, we briefly
review the holographic DE model with new IR cut-off. In section III,
we perform the cosmic observation constraint on the holographic DE
model. The last section is the conclusion.

\section{Review of Holographic Dark Energy Model with New Infrared cut-off}
In this section, we give a brief review on the general formula in
the holographic DE model with new IR cut-off. With a
Friedmann-Robertson-Walker (FRW) metric
\begin{eqnarray}
&&ds^2=-dt^2+a^2(t)[\frac{dr^2}{1-kr^2}+r^2(d{\theta}^2+\sin^2{\theta}d{\phi}^2)],
\end{eqnarray}
the Einstein field equation can be written as
\begin{eqnarray}
&&H^2=\frac{1}{3M_{pl}^2}\sum_{i}\rho_i, \\
&&\frac{\ddot{a}}{a}=-\frac{1}{6M_{pl}^2}\sum_{i}(\rho_i+3P_i),
\end{eqnarray}
where $H$ is the Hubble function, and $\rho_i$ and $P_i$ are the
energy density and the pressure of a general piece of matter, and
their subscripts $i$ denote $m$, $de$ and $k$, which respectively
correspond to matter component, the holographic DE with new infrared
cut-off and the curvature part of space. Here the matter component
includes the cold dark matter and the baryon matter, i.e.
\begin{eqnarray}
&&\rho_m=\rho_{cdm}+\rho_b, P_m=0.
\end{eqnarray}
The parameter $k=1, 0, -1$ denote the closed, flat and open
geometries, respectively. The effective energy density and the
effective pressure of the curvature part are
\begin{eqnarray}
&&\rho_k=-\frac{3M_{pl}^2k}{a^2}, \\
&&P_k=-\rho_k-\frac{\dot{\rho}_k}{3H}.
\end{eqnarray}

As suggested by Granda and Oliveros in paper \cite{ref:holo4}, the
energy density of the holographic DE with new IR cut-off is given as
\begin{eqnarray}
&&\rho_{de}=3M_{pl}^2(\alpha H^2+\beta \dot{H}),
\end{eqnarray}
where $\alpha$ and $\beta$ are the dimensionless parameters in
holographic DE model with new IR cut-off, which are regarded as
independent of each other. In this paper, a dot denotes a derivative
with respect to the cosmic time $t$. After changing the variable
from the cosmic time $t$ to $x=\ln a$, we can rewritten the Eq. (4)
as
\begin{eqnarray}
&&H^2=\frac{1}{3M_{pl}^2}\rho_{m0}e^{-3x}-ke^{-2x}+\alpha
H^2+\frac{1}{2}\beta\frac{dH^2}{dx}.
\end{eqnarray}
With the help of the definitions as follows:
\begin{eqnarray}
&&E=
\frac{H}{H_0},\Omega_{m0}=\frac{\rho_{m0}}{3M_{pl}^2H_0^2},\Omega_k=-\frac{k}{H_0^2},
\end{eqnarray}
the Eq. (10) can be ulteriorly rewritten as
\begin{eqnarray}
&&E^2=\Omega_ke^{-2x}+\Omega_{m0}e^{-3x}+\alpha
E^2+\frac{1}{2}\beta\frac{dE^2}{dx}.
\end{eqnarray}
Solving this first order differential equation about $E^2$, we can
obtain
\begin{eqnarray}
E^2&=&\Omega_ke^{-2x}+\Omega_{m0}e^{-3x}+\frac{2\alpha-3\beta}{3\beta-2\alpha+2}\Omega_{m0}e^{-3x}+
\frac{\alpha-\beta}{\beta-\alpha+1}\Omega_ke^{-2x}+f_0e^{-\frac{2(\alpha-1)}{\beta}x} \nonumber\\
&=&\Omega_ke^{-2x}+\Omega_{m0}e^{-3x}+\Omega_{de}(x),
\end{eqnarray}
where $f_0$ is the integral constant and can be derived from the
initial condition $E_0=1$, which is
$f_0=1-\frac{1}{\beta-\alpha+1}\Omega_k-\frac{2}{3\beta-2\alpha+2}\Omega_{m0}$,
and $\Omega_{de}(x)$ is the dimensionless energy density of the
holographic DE with new IR cut-off:
\begin{eqnarray}
&&\Omega_{de}(x)=\frac{2\alpha-3\beta}{3\beta-2\alpha+2}\Omega_{m0}e^{-3x}+\frac{\alpha-\beta}{\beta-\alpha+1}
\Omega_ke^{-2x}+(1-\frac{1}{\beta-\alpha+1}\Omega_k-\frac{2}{3\beta-2\alpha+2}\Omega_{m0})e^{-\frac{2(\alpha-1)}{\beta}x},
\end{eqnarray}
Then, combining the above definition of the dimensionless energy
density of the holographic DE with its conservation equation, we can
obtain the EOS of the holographic DE with new IR cut-off
\begin{eqnarray}
&&w_{de}(z)=-1+\frac{1+z}{3}\frac{d\ln{\Omega_{de}}}{dz}.
\end{eqnarray}
In addition, we shall investigate the evolution of the deceleration
parameter. Combining Eqs. (4) and (5) with the definition of the
deceleration parameter, we can get
\begin{eqnarray}
q(z)&&{=}\frac{1}{2}+\frac{3}{2}\frac{\sum_{i}\tilde{P_i}}{\sum_{i}\tilde{\rho_i}} \nonumber\\
&&{=}\frac{1}{2}+\frac{3}{2}\frac{\sum_{i}\tilde{P_i}}{\Omega_{m0}(1+z)^3+\Omega_k(1+z)^2+\Omega_{de}(z)},
\end{eqnarray}
where we have utilized the definitions of
$\tilde{P_i}=\frac{P_i}{3M_{pl}^2H_0^2}$ and
$\tilde{\rho_i}=\frac{\rho_i}{3M_{pl}^2H_0^2}$. According to the
energy conservation equation, we have
\begin{eqnarray}
&&\tilde{P_i}=-\tilde{\rho_i}-\frac{1}{3}\frac{d\tilde{\rho_i}}{dx}.
\end{eqnarray}

In the review above, it is direct and natural to consider the
parameters $\beta\neq 0$ and $\alpha\neq 1$ when we solve the
differential Eq. (12). Next, we discuss three special cases when the
denominators in Eq. (13) equal zero as follows:

{\bf Case 1:} $3\beta-2\alpha+2=0$ and $\beta-\alpha+1=0$

 In this case, we obtain $\beta=0$ and $\alpha=1$. Now the energy density of the
holographic DE is $\rho_{de}=3M_{pl}^2H^2$. Compared to the
Friedmann equation, it is found that the DE density is not
consistent with the current observations of $70\%$ exotic component
in the universe.

{\bf Case 2:} $3\beta-2\alpha+2\neq 0$ and $\beta-\alpha+1=0$

 From the latter equation, we can get $\beta=\alpha-1$. Then the energy density of the holographic DE is
 $\rho_{de}=3M_{pl}^2(\frac{\alpha-1}{2}\frac{dH^2}{dx}+\alpha H^2)$. In this case, the solution of the first order differential equation is
\begin{eqnarray}
E^2&&{=}f_0e^{-2x}+\frac{2}{\alpha-1}\Omega_{m0}e^{-3x}-\frac{2}{\alpha-1}x\Omega_ke^{-2x} \nonumber\\
&&{=}\Omega_ke^{-2x}+\Omega_{m0}e^{-3x}+\Omega_{de}(x),
\end{eqnarray}
where we have set the integral constant $f_0=\Omega_k$ and defined
$\Omega_{de}(x)=\frac{3-\alpha}{\alpha-1}\Omega_{m0}e^{-3x}-
\frac{2}{\alpha-1}x\Omega_ke^{-2x}$ as the dimensionless energy
density of the holographic DE. Considering the current value of the
dimensionless DE density, we can obtain the value of the only model
parameter
$\alpha=\frac{3\Omega_{m0}+\Omega_{de0}}{\Omega_{m0}+\Omega_{de0}}$.
So the case is a viable DE model.

{\bf Case 3:} $3\beta-2\alpha+2=0$ and $\beta-\alpha+1\neq 0$

From the former equation, we can obtain
$\beta=\frac{2(\alpha-1)}{3}$. Then the energy density of the
holographic DE is
$\rho_{de}=3M_{pl}^2(\frac{\alpha-1}{3}\frac{dH^2}{dx}+\alpha H^2)$.
In this case, the solution of the first order differential equation
is
\begin{eqnarray}
E^2&&{=}f_0e^{-3x}-\frac{3}{\alpha-1}x\Omega_{m0}e^{-3x}-\frac{3}{\alpha-1}\Omega_ke^{-2x} \nonumber\\
&&{=}\Omega_{m0}e^{-3x}+\Omega_ke^{-2x}+\Omega_{de}(x),
\end{eqnarray}
where we have taken the integral constant $f_0=\Omega_{m0}$ and
defined $\Omega_{de}(x)=-\frac{3}{\alpha-1}x\Omega_{m0}e^{-3x}-
\frac{2+\alpha}{\alpha-1}\Omega_ke^{-2x}$ as the dimensionless
energy density of the holographic DE. Considering the current value
of the dimensionless DE density, we can get the value of the only
model parameter
$\alpha=\frac{\Omega_{de0}-2\Omega_k}{\Omega_{de0}+\Omega_k}$. So
the case with the non-flat background geometry is a viable dark
energy model. The result in the flat case is the same as {\bf Case
1}.

As far as the {\bf Case 2} and {\bf Case 3} discussed above are
concerned, the model parameters $\alpha$ and $\beta$ are reliant on
each other. In this paper, we consider the generalized case with
independent model parameters. In the next section, according to the
combined observational data, we restrict the basic cosmological
parameters and the model parameters, all of which are independent on
each other.
\section{Method and results}

\begin{table}
\begin{center}
\begin{tabular}{cc|   cc   cc   cc   cc}
\hline\hline parameters && flat holographic  &  & not-flat
holographic & & flat $\Lambda \textmd{CDM}$  &  & not-flat $\Lambda
\textmd{CDM}$ &
\\ \hline
$\Omega_{b}h^2$ &&$0.0233^{+0.0009 +0.0013}_{-0.0009 -0.0014}$ &
                 &$0.0228^{+0.0010 +0.0014}_{-0.0010 -0.0014}$ &
                 &$0.0228^{+0.0007 +0.0011}_{-0.0007 -0.0011}$ &
                 &$0.0228^{+0.0007 +0.0011}_{-0.0008 -0.0013}$ & \\
$\Omega_{k}$    && - &
                 &$0.0305^{+0.0092 +0.0140}_{-0.0134 -0.0176}$ &
                 & - &
                 &$-0.0013^{+0.0070 +0.0103}_{-0.0076 -0.0108}$ &\\
$\alpha$        &&$0.8502^{+0.0984 +0.1299}_{-0.0875 -0.1064}$ &
                 &$0.8824^{+0.2180 +0.2213}_{-0.1163 -0.1378}$ &
                 & - &
                 & - &\\
$\beta$         &&$0.4817^{+0.0842 +0.1176}_{-0.0773 -0.0955}$ &
                 &$0.5016^{+0.0973 +0.1247}_{-0.0871 -0.1102}$ &
                 & - &
                 & - &\\
$\Omega_{de0}/ \Omega_{\Lambda0}$  &&$0.7287^{+0.0296
+0.0432}_{-0.0294 -0.0429}$ & &$0.6934^{+0.0364 +0.0495}_{-0.0304
-0.0413}$ &
                 &$0.7220^{+0.0177 +0.0273}_{-0.0185 -0.0315}$ &
                 &$0.7258^{+0.0222 +0.0317}_{-0.0268 -0.0407}$ &\\
$\Omega_{m0}$   &&$0.2713^{+0.0294 +0.0429}_{-0.0296 -0.0432}$ &
                 &$0.2762^{+0.0278 +0.0402}_{-0.0320 -0.0412}$ &
                 &$0.2780^{+0.0185 +0.0315}_{-0.0177 -0.0273}$ &
                 &$0.2755^{+0.0239 +0.0367}_{-0.0186 -0.0274}$ &\\
$H_0$           &&$66.35^{+2.38 +3.35}_{-2.14 -3.07}$ &
                 &$70.20^{+3.03 +3.58}_{-3.17 -4.00}$ &
                 &$70.11^{+1.44 +2.35}_{-1.34 -2.24}$ &
                 &$70.04^{+1.72 +2.81}_{-2.20 -2.91}$ &       \\
 \hline $\chi^{2}/dof$    && $1.23849$ &  & $1.15996$ &     & $1.15339$ &  & $1.15589$&   \\
\hline\hline
\end{tabular}
\caption{The data fitting results of the cosmological parameters and
the model parameters with $1\sigma$, $2\sigma$ regions in the flat
and non-flat holographic DE model with new infrared cut-off and
$\Lambda \textmd{CDM}$ model (flat case and non-flat case), where
the combined observational data from SN $397$, BAO and CMB and OHD
are used.}\label{tab:results}
\end{center}
\end{table}

\begin{figure}[!htbp]
\includegraphics[width=20cm,height=14cm]{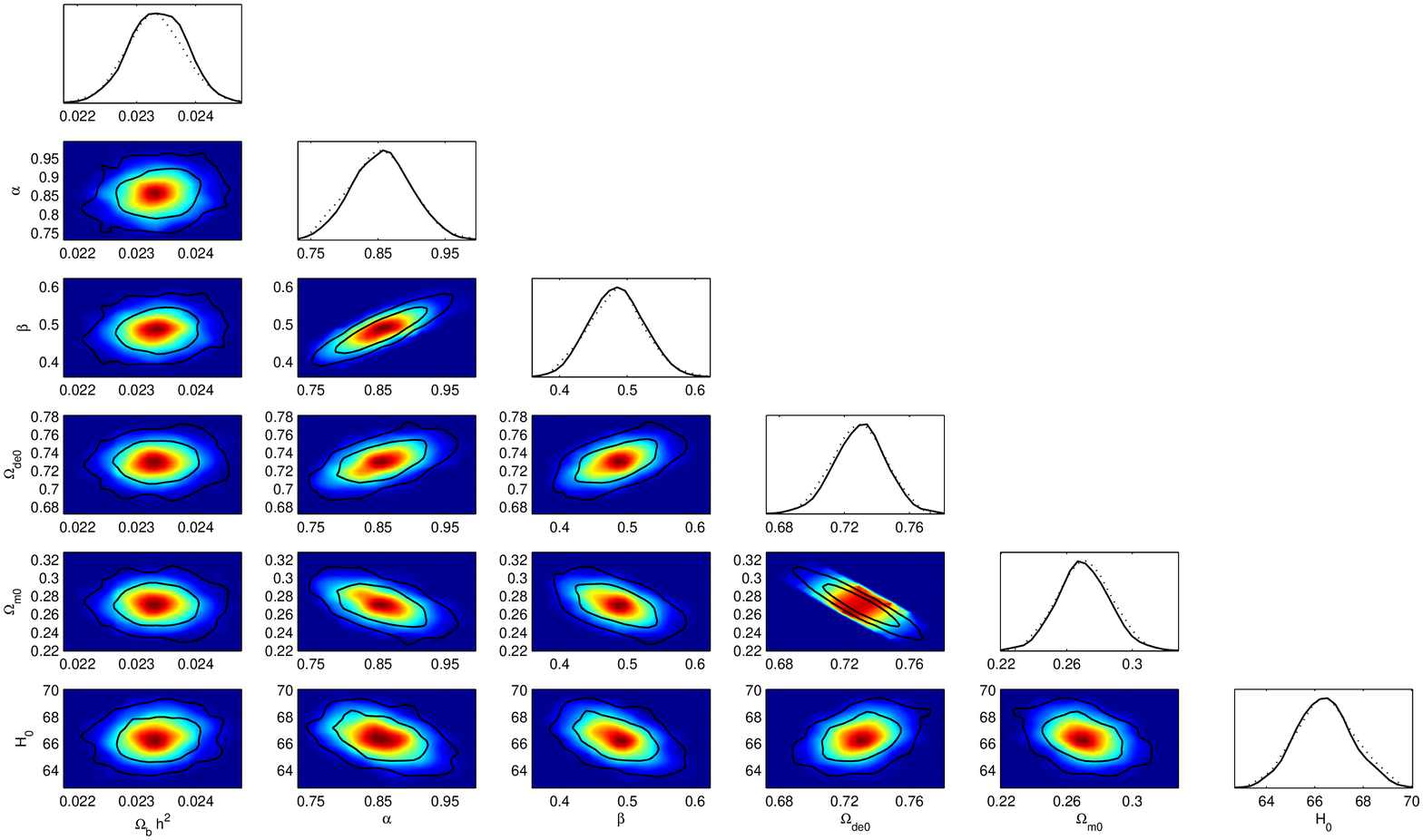}
  \caption{1-D constraints on individual parameters ($\Omega_bh^2, \alpha, \beta, \Omega_{de0}, \Omega_{m0},
  H_0$) and 2-D contours on these parameters with $1\sigma, 2\sigma$ errors between each other using the combination of the observational
  data from SN $397$, BAO, CMB and OHD in the flat holographic DE model with new IR cut-off. Dotted lines in the 1-D plots
  show the mean likelihood of the samples and the solid lines are marginalized
  probabilities for the parameters in the flat holographic DE model with new IR cut-off \cite{ref:MCMC}.
 }\label{fig:NRicciflat}
\end{figure}

\begin{figure}[!htbp]
\includegraphics[width=20cm,height=14cm]{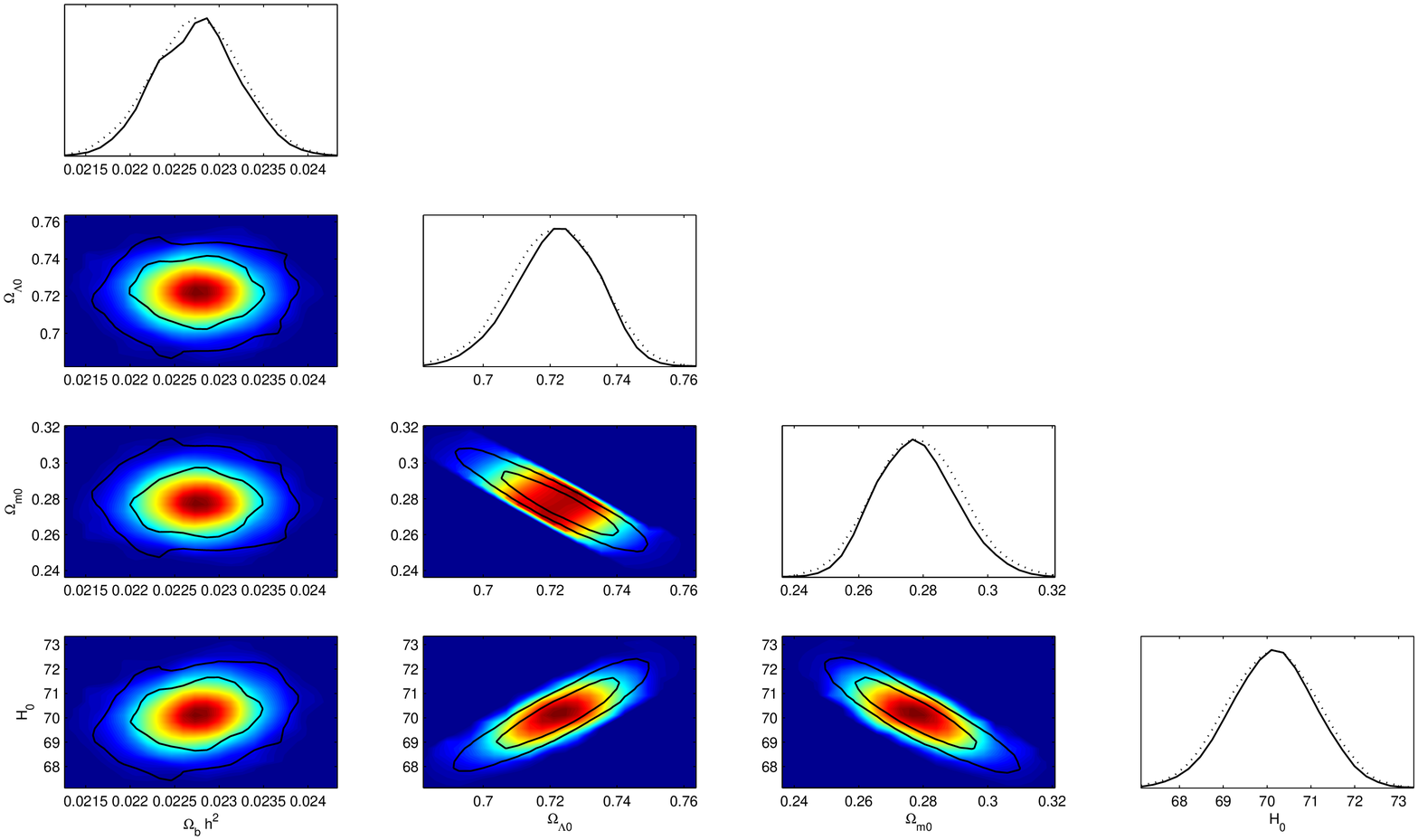}
  \caption{1-D constraints on individual parameters ($\Omega_bh^2, \Omega_{\Lambda0}, \Omega_{m0},
  H_0$) and 2-D contours on these parameters with $1\sigma, 2\sigma$ errors between each other using the combination of the observational
  data from SN $397$, BAO, CMB and OHD in the flat $\Lambda \textmd{CDM}$ model. Dotted lines in the 1-D plots
  show the mean likelihood of the samples and the solid lines are marginalized
  probabilities for the parameters in the flat $\Lambda \textmd{CDM}$ model \cite{ref:MCMC}.
 }\label{fig:LCDMflat}
\end{figure}

\begin{figure}[!htbp]
\includegraphics[width=20cm,height=14cm]{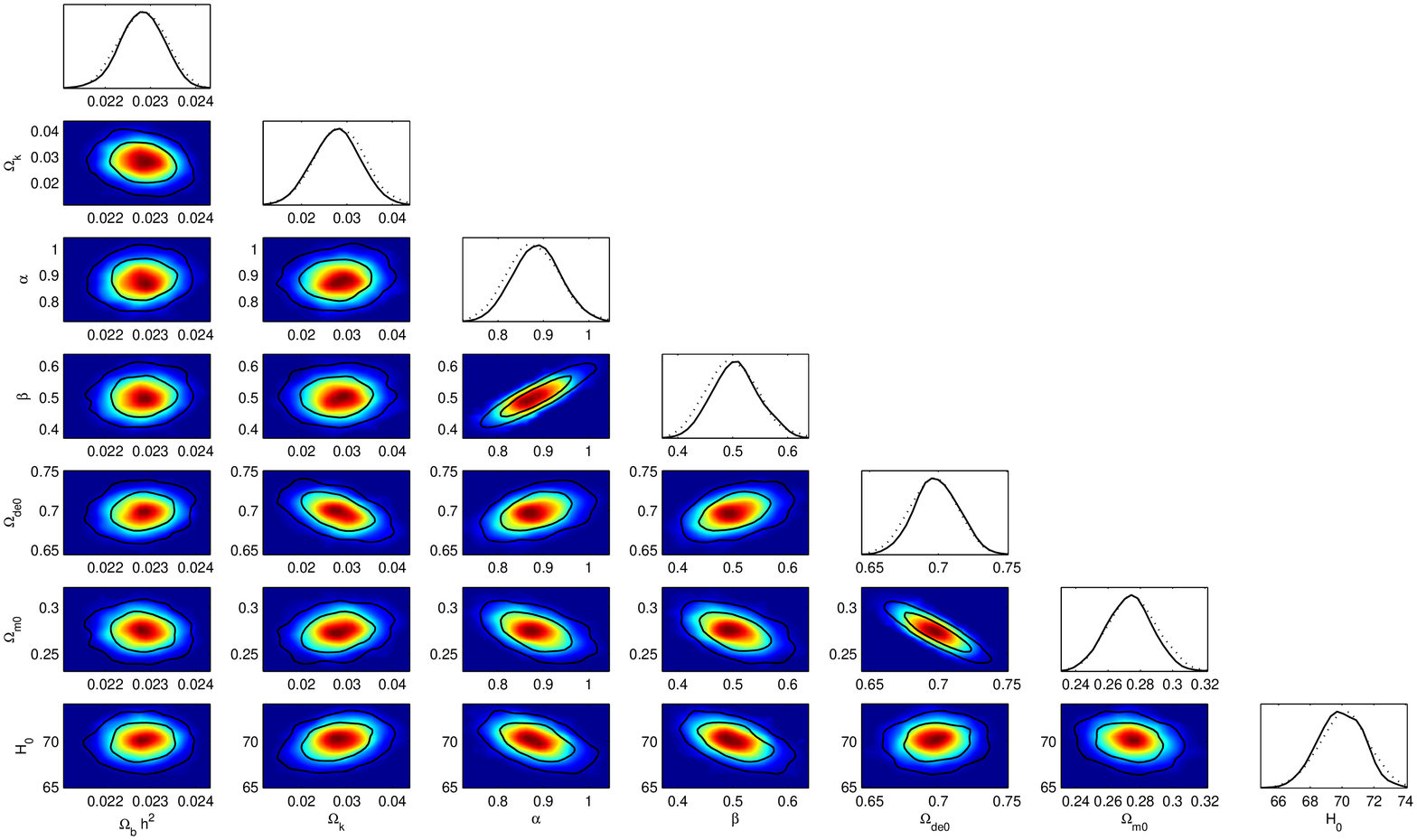}
  \caption{1-D constraints on individual parameters ($\Omega_bh^2, \Omega_k, \alpha, \beta, \Omega_{de0}, \Omega_{m0},
  H_0$) and 2-D contours on these parameters with $1\sigma, 2\sigma$ errors between each other using the combination of the observational
  data from SN $397$, BAO, CMB and OHD in the non-flat holographic DE model with new IR cut-off. Dotted lines in the 1-D plots
  show the mean likelihood of the samples and the solid lines are marginalized
  probabilities for the parameters in the non-flat holographic DE model with new IR cut-off \cite{ref:MCMC}.
 }\label{fig:NRiccinonflat}
\end{figure}

\begin{figure}[!htbp]
\includegraphics[width=20cm,height=14cm]{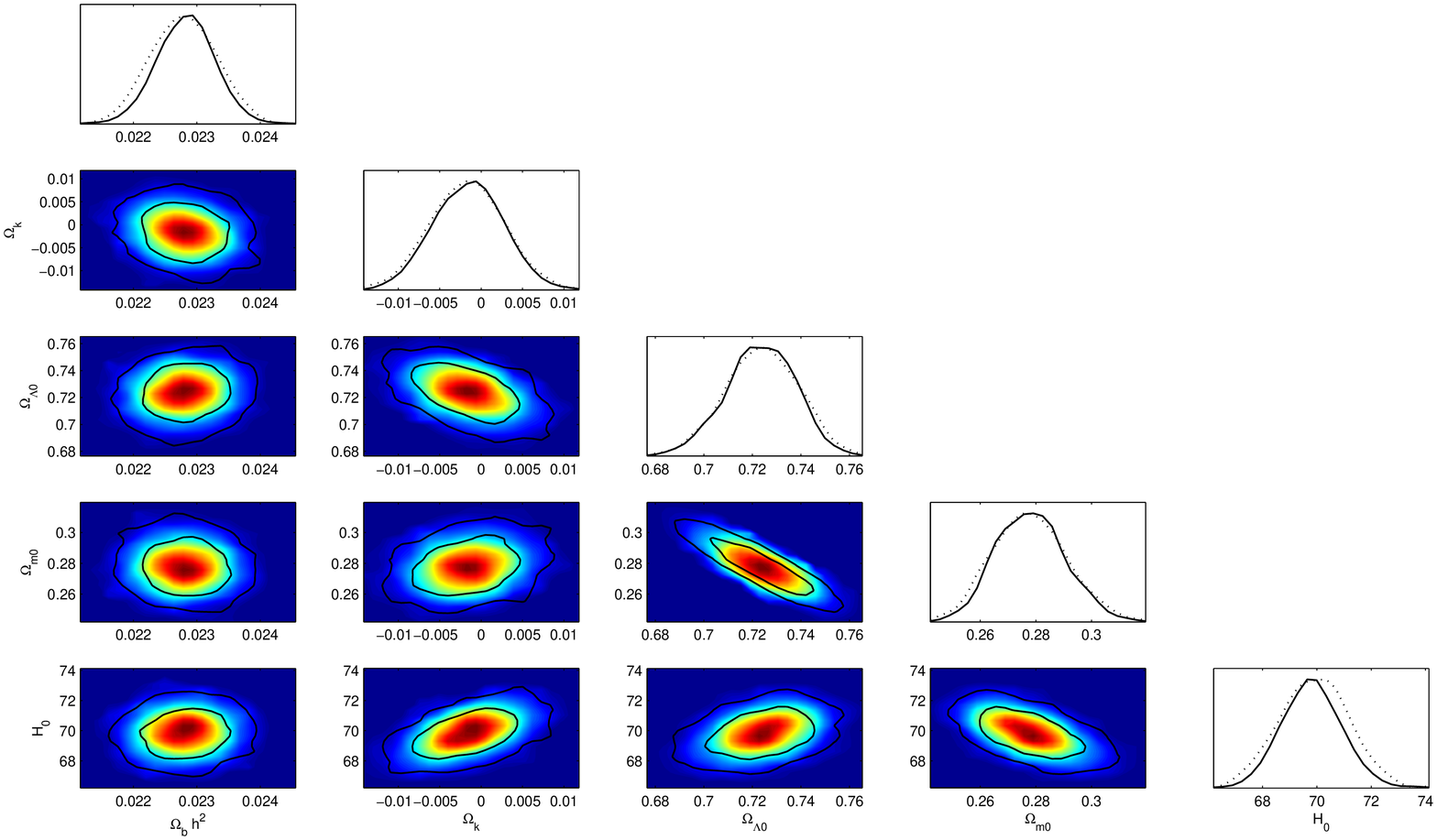}
  \caption{1-D constraints on individual parameters ($\Omega_bh^2, \Omega_k, \Omega_{\Lambda0}, \Omega_{m0},
  H_0$) and 2-D contours on these parameters with $1\sigma, 2\sigma$ errors between each other using the combination of the observational
  data from SN $397$, BAO, CMB and OHD in the non-flat $\Lambda \textmd{CDM}$ model. Dotted lines in the 1-D plots
  show the mean likelihood of the samples and the solid lines are marginalized
  probabilities for the parameters in the non-flat $\Lambda \textmd{CDM}$ model \cite{ref:MCMC}.
 }\label{fig:LCDMnonflat}
\end{figure}
In this section, we present the method and the data we have used. In
our analysis, we perform a global fitting on determining the
cosmological parameters using the Markov Chain Monte Carlo (MCMC)
method. Since the computational requirements of MCMC procedures are
insensitive to the dimensionality of the parameter space, we can
expand the dimension of the parameter series, comparing with the
traditional Maximum Likelihood (ML) method. The MCMC method is based
on the publicly available {\bf CosmoMC} package \cite{ref:MCMC},
which has been modified to include the new parameters $\alpha$ and
$\beta$ with having taken the weak priors as $\alpha\in[0.5,1.5]$
and $\beta\in[0.1,1.0]$. Besides the two independent model
parameters, the basic cosmological parameters are also varying with
top-hat priors: the physical baryon density
$\Omega_{b}h^2\in[0.005,0.9]$, the dark matter energy density
$\Omega_{c}h^2\in[0.01,0.99]$, and in the non-flat case, the
additional parameter $\Omega_k\in[-0.1,0.1]$. In addition, we obtain
three derived parameters $\Omega_{de0}$, $\Omega_{m0}$ and the
Hubble constant $H_0$ from the basic cosmological parameters.

In our calculations, we have taken the total likelihood $L\propto
e^{-\chi^2/2}$ to be the product of the separate likelihoods of SN,
BAO, CMB and OHD. Then the $\chi^2$ is
\begin{eqnarray}
\chi^2=\chi^2_{SN}+\chi^2_{BAO}+\chi^2_{CMB}+\chi^2_{OHD}.
\end{eqnarray}
The expressions of $\chi^2$s and datasets used in our paper are
presented in Appendix \ref{app}.

\begin{figure}[!htbp]
\includegraphics[width=8cm]{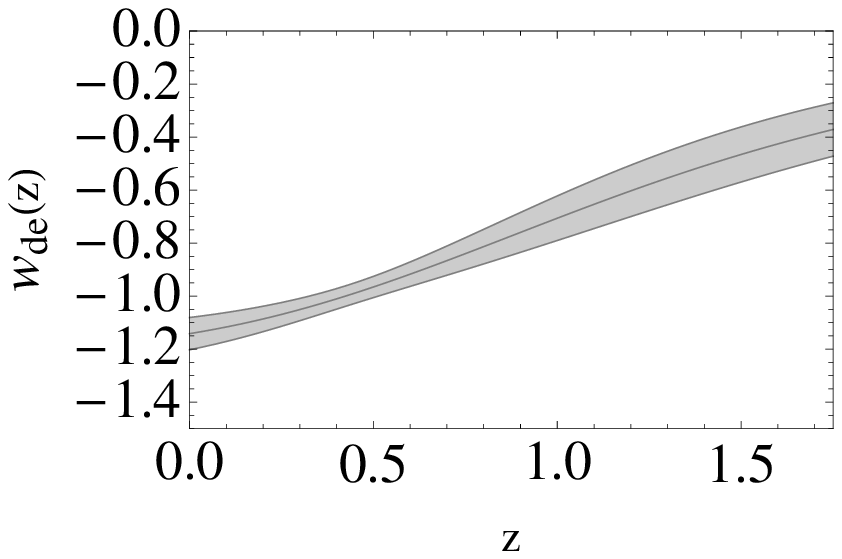}
 ~~~\includegraphics[width=8cm]{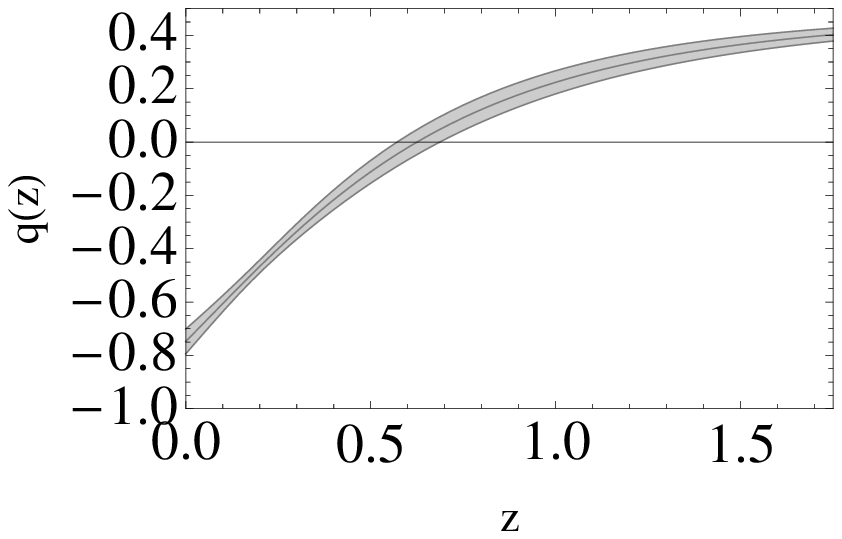}\\
  \caption{The evolutions of the EOS (the left panel) and the deceleration parameter (the right panel) in the flat
  holographic DE model with new IR
  cut-off with respect to the redshift $z$, where we have used the best fittings of the cosmological
  parameters and model parameters with $1\sigma$ errors.
 }\label{fig:wqflat}
\end{figure}

The best fit values of the cosmological parameters and the model
parameters with $1\sigma, 2\sigma$ errors in holographic DE model
with new IR cut-off and the $\Lambda \textmd{CDM}$ model for the
flat case and the non-flat case are listed in Table
\ref{tab:results}. We calculate the values of $\chi^2/dof$, where
$dof$ is the compact notation of the number of degrees of freedom
and equals the number of observational data points minus the number
of free parameters. It is found that the values of $\chi^2/dof$
exhibit a significant difference between the flat case and the
non-flat case in the holographic DE model with the new IR cut-off.
In this two instances, it is seen that the non-flat holographic DE
model with a smaller value of $\chi^2/dof$ is much supported by the
current observations. Subsequently, comparing the value of
$\chi^2/dof$ in the non-flat holographic DE model with those in the
$\Lambda$CDM models, we find the differences are not obvious. From
the minor differences among the values of $\chi^2/dof$, we can
conclude that the current combined datasets do not really favor the
holographic DE model with the new IR cut-off over the concordance
model. It is seen that the current observations support the flat
concordance model with the smallest value of $\chi^2/dof$ the best.
In Fig. \ref{fig:NRicciflat}, \ref{fig:LCDMflat}, we show one
dimensional probability distribution of each parameter and two
dimensional plots for parameters between each other in the flat
holographic DE model with new IR cut-off and the flat $\Lambda
\textmd{CDM}$ model. The corresponding plots in the non-flat
holographic DE model with new IR cut-off and the non-flat $\Lambda
\textmd{CDM}$ model are presented in Fig.
\ref{fig:NRiccinonflat},\ref{fig:LCDMnonflat}.

\begin{figure}[!htbp]
\includegraphics[width=8cm]{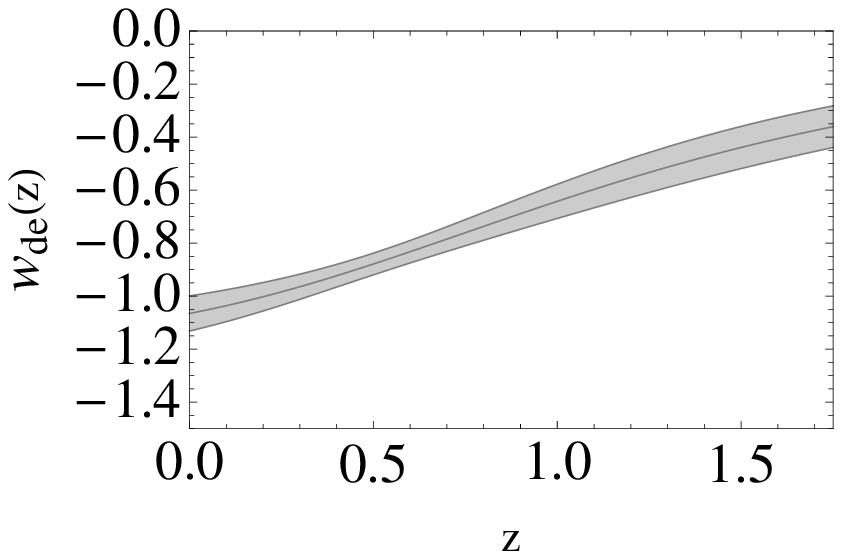}
 ~~~\includegraphics[width=8cm]{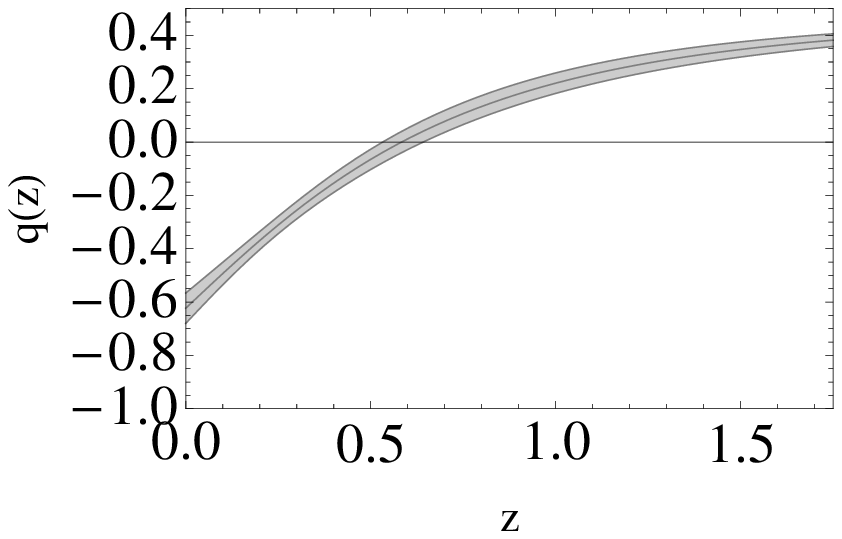}\\
  \caption{The evolutions of the EOS (the left panel) and the deceleration parameter (the right panel) in the non-flat
  holographic DE model with new IR
  cut-off with respect to the redshift $z$, where we have used the best fittings of the cosmological
  parameters and model parameters with $1\sigma$ errors.
 }\label{fig:wqnonflat}
\end{figure}

Then, we investigate the evolutions of dark energy EOS and
deceleration parameter in the holographic DE model. We consider the
propagation of the errors for $w(z)$ and $q(z)$ by the Fisher matrix
analysis. The errors are evaluated by using the covariance matrix
$C_{ij}$ of the fitting parameters \cite{ref:wqerror1,ref:wqerror2},
which is the inverse of the Fisher matrix and given by
\begin{eqnarray}
&&(C_{ij})^{-1}=-\frac{\partial^2\ln
L}{\partial\theta_i\partial\theta_j},
\end{eqnarray}
where $\theta$ is a set of parameters, and $\ln L$ is the
logarithmic likelihood function. The errors on a function
$f=f(\theta)$ in terms of the variables $\theta$ are given by
\cite{ref:wqerror2,ref:wqerror3}
\begin{eqnarray}
&&\sigma_{f}^2=\sum_i^n\left(\frac{\partial
f}{\partial\theta_i}\right)^2C_{ii}+2\sum_i^n\sum_{j=i+1}^n\left(\frac{\partial
f}{\partial\theta_i}\right)\left(\frac{\partial
f}{\partial\theta_j}\right)C_{ij},
\end{eqnarray}
where $n$ is the number of parameters. Here, $f$ will be dark energy
EOS $w(z;\theta_i)$ or deceleration parameter $q(z;\theta_i)$. The
parameters $\theta_i$ respectively represent
($\Omega_ch^2,\Omega_bh^2,\alpha,\beta$) for the flat case and
($\Omega_ch^2,\Omega_bh^2,\Omega_k,\alpha,\beta$) for the non-flat
case.
As shown in Fig. \ref{fig:wqflat} (flat case) and Fig.
\ref{fig:wqnonflat} (non-flat case), we plot the evolutions of
$w(z)$ and $q(z)$ with errors by
\begin{eqnarray}
&&w_{1\sigma}(z)=w(z)|_{\theta=\bar{\theta}}\pm\sigma_w, \\
&&q_{1\sigma}(z)=q(z)|_{\theta=\bar{\theta}}\pm\sigma_q,
\end{eqnarray}
where $\bar{\theta}$ are the best fit values of the constraint
parameters.

In Fig. \ref{fig:wqflat} and Fig. \ref{fig:wqnonflat}, it is found
that the combined observational data provide a fairly tight
constraint on the holographic DE model with new IR cut-off. From the
left panel in Fig. \ref{fig:wqflat}, it is seen that the EOS
$w_{de}(z)$ with the best fit values can cross the boundary $-1$ and
its current value is $w_{de}(z=0)=-0.1414$. From the right panel in
Fig. \ref{fig:wqflat}, the value of the current deceleration
parameter is given by $q_0=-0.7476\pm0.0466$. In Fig.
\ref{fig:wqnonflat}, we get the current value of the dark energy EOS
in the non-flat case is $w_{de}(z=0)=-1.0653<-1$, i.e. it is also
phantom-like. The present value of the deceleration parameter is
$q_0=-0.6231\pm0.0569$.

\section{Conclusion}
In summary, in this paper we have performed a global fitting on the
parameters in the holographic DE model with new IR cut-off for the
flat case and the non-flat case, using a combined cosmic
observations from type Ia supernovae, baryon acoustic oscillations,
Cosmic Microwave Background and the observational Hubble data. The
same constraints are performed on the flat and non-flat concordance
models by using the same combined datasets. According to the Markov
Chain Monte Carlo (MCMC) analysis, it is shown that the best fitting
values of the model parameters ($\alpha,\beta$) in the flat
holographic DE model with new IR cut-off tend to be smaller than
those in the non-flat case. In the holographic DE models, the
non-flat case with a smaller value of $\chi^2/dof$ is much supported
by the observations. In the non-flat cases, we have obtained the
constraint values of the curvature terms $\Omega_k=0.0305^{+0.0092
+0.0140}_{-0.0134 -0.0176}$ for the holographic DE model with new IR
cut-off and $\Omega_k=-0.0013^{+0.0070 +0.0103}_{-0.0076 -0.0108}$
for the concordance model. These results indicate the two kinds of
the non-flat background geometries in the two models. Then by using
the best fit parameters, we plot the evolutions of the dark energy
EOS and deceleration parameter with errors. From Fig.
\ref{fig:wqflat} and Fig. \ref{fig:wqnonflat}, it is found that the
EOS of the holographic DE with new IR cut-off can cross the phantom
divide $-1$, respectively with the current best values
$w_{de0}=-1.1414$ (flat case) and $w_{de0}=-1.0653$ (non-flat case).
Comparing the flat and non-flat holographic DE models with the
corresponding cases in the $\Lambda \textmd{CDM}$ model, we can find
that the current combined observations do not favor the holographic
DE model with new IR cut-off over the $\Lambda \textmd{CDM}$ model.
\section*{Acknowledgments}
The data fitting is based on the publicly available {\bf CosmoMC}
package a Markov Chain Monte Carlo (MCMC) code. This work is
supported by the National Natural Science Foundation of China (Grant
No 10703001), and Specialized Research Fund for the Doctoral Program
of Higher Education (Grant No 20070141034).

\appendix

\section{Cosmological Constraints Methods and Dataset}\label{app}
\subsection{Type Ia Supernovae constraints}

We use the SN Ia Constitution dataset, which includes $397$ SN Ia
\cite{ref:Condata}. The 90 SN Ia from CfA3 sample with low redshifts
are added to 307 SN Ia Union sample \cite{ref:Kowalski}. The CfA3
sample increases the number of the nearby SN Ia and reduces the
statistical uncertainties. Following
\cite{ref:smallomega,ref:POLARSKI}, one can obtain the corresponding
constraint by fitting the distance modulus $\mu(z)$ as
\begin{equation}
\mu_{th}(z)=5\log_{10}[D_{L}(z)]+\mu_{0}.
\end{equation}
In this expression $D_{L}(z)$ is the Hubble-free luminosity distance
$H_0 d_L(z)/c$, with $H_0$ the Hubble constant, defined through the
re-normalized quantity $h$ as $H_0 =100 h~{\rm km ~s}^{-1} {\rm
Mpc}^{-1}$,
 and
\begin{eqnarray}
d_L(z)&=&\frac{c(1+z)}{\sqrt{|\Omega_k|}}\textmd{sinn}[\sqrt{|\Omega_k|}\int_0^z\frac{dz'}{H(z')}],\\
\mu_0&\equiv&42.38-5\log_{10}h,
\end{eqnarray}
where $\textmd{sinnn}(\sqrt{|\Omega_k|}x)$ respectively denotes
$\sin(\sqrt{|\Omega_k|}x)$, $\sqrt{|\Omega_k|}x$,
$\sinh(\sqrt{|\Omega_k|}x)$ for $\Omega_k<0$, $\Omega_k=0$ and
$\Omega_k>0$.
 Additionally, the observed distance moduli $\mu_{obs}(z_i)$ of SN
Ia at $z_i$ is
\begin{equation}
\mu_{obs}(z_i) = m_{obs}(z_i)-M,
\end{equation}
where $M$ is their absolute magnitudes.

For the SN Ia dataset, the best fit values of the parameters $p_s$
can be determined by a likelihood analysis, based on the calculation
of
\begin{eqnarray}
\chi^2(p_s,M^{\prime})\equiv \sum_{SN}\frac{\left\{
\mu_{obs}(z_i)-\mu_{th}(p_s,z_i)\right\}^2} {\sigma_i^2} \ \ \ \ \ \ \ \ \ \  \ \ \ \ \ \ \nonumber\\
=\sum_{SN}\frac{\left\{ 5 \log_{10}[D_L(p_s,z_i)] - m_{obs}(z_i) +
M^{\prime} \right\}^2} {\sigma_i^2}, \ \ \ \ \label{eq:chi2}
\end{eqnarray}
where $M^{\prime}\equiv\mu_0+M$ is a nuisance parameter which
includes the absolute magnitude and the parameter $h$. The nuisance
parameter $M^{\prime}$ can be marginalized over analytically
\cite{ref:SNchi2} as
\begin{equation}
\bar{\chi}^2(p_s) = -2 \ln \int_{-\infty}^{+\infty}\exp \left[
-\frac{1}{2} \chi^2(p_s,M^{\prime}) \right] dM^{\prime},\nonumber
\label{eq:chi2marg}
\end{equation}
to obtain
\begin{equation}
\bar{\chi}^2 =  A - \frac{B^2}{C} + \ln \left( \frac{C}{2\pi}\right)
, \label{eq:chi2mar}
\end{equation}
with
\begin{eqnarray}
&&A=\sum_{SN} \frac {\left\{5\log_{10}
[D_L(p_s,z_i)]-m_{obs}(z_i)\right\}^2}{\sigma_i^2},\nonumber\\
&& B=\sum_{SN} \frac {5
\log_{10}[D_L(p_s,z_i)]-m_{obs}(z_i)}{\sigma_i^2},\nonumber
\\
&& C=\sum_{SN} \frac {1}{\sigma_i^2}\nonumber.
\end{eqnarray}
Relation (\ref{eq:chi2}) has a minimum at the nuisance parameter
value $M^{\prime}=B/C$, which contains information of the values of
$h$ and $M$. Therefore, one can extract the values of $h$ and $M$
provided one get the knowledge of one of them. Finally, it is noted
that the expression
\begin{equation}
\chi^2_{SN}(p_s,B/C)=A-(B^2/C),\label{eq:chi2SN}\nonumber
\end{equation}
which coincides to (\ref{eq:chi2mar}) up to a constant, is often
used in the likelihood analysis
\cite{ref:smallomega,ref:JCAPXU,ref:SNchi2}, and thus in this case
the results will not be affected by a flat $M^{\prime}$
distribution.

\subsection{Baryon Acoustic Oscillation constraints}

 The Baryon Acoustic Oscillations are detected in the clustering of the
combined 2dFGRS and SDSS main galaxy samples, and measure the
distance-redshift relation at $z = 0.2$. Additionally, Baryon
Acoustic Oscillations in the clustering of the SDSS luminous red
galaxies measure the distance-redshift relation at $z = 0.35$. The
observed scale of the BAO calculated from these samples, as well as
from the combined samples, are jointly analyzed using estimates of
the correlated errors to constrain the form of the distance measure
$D_V(z)$ \cite{ref:Okumura2007,ref:Percival2,ref:Percival3}
\begin{equation}
D_V(z)=\left[(1+z)^2 D^2_A(z) \frac{cz}{H(z)}\right]^{1/3}.
\label{eq:DV}
\end{equation}
In this expression  $D_A(z)$ is the proper (not comoving) angular
diameter distance, which has the following relation with $d_{L}(z)$
\begin{equation}
D_A(z)=\frac{d_{L}(z)}{(1+z)^2}.
\end{equation}
The peak positions of the BAO depend on the ratio of $D_V(z)$ to the
sound horizon size at the drag epoch (where baryons were released
from photons) $z_d$, which can be obtained by using a fitting
formula \cite{ref:Eisenstein}:
\begin{eqnarray}
&&z_d=\frac{1291(\Omega_mh^2)^{-0.419}}{1+0.659(\Omega_mh^2)^{0.828}}[1+b_1(\Omega_bh^2)^{b_2}],
\end{eqnarray}
with
\begin{eqnarray}
&&b_1=0.313(\Omega_mh^2)^{-0.419}[1+0.607(\Omega_mh^2)^{0.674}], \\
&&b_2=0.238(\Omega_mh^2)^{0.223}.
\end{eqnarray}
In this paper, we use the data of $r_s(z_d)/D_V(z)$ extracted from
the Sloan Digitial Sky Survey (SDSS) and the Two Degree Field Galaxy
Redshift Survey (2dFGRS) \cite{ref:Percival3}, which are listed in
Table \ref{baodata}, where $r_s(z)$ is the comoving sound horizon
size
\begin{eqnarray}
r_s(z)&&{=}c\int_0^t\frac{c_sdt}{a}=c\int_0^a\frac{c_sda}{a^2H}=c\int_z^\infty
dz\frac{c_s}{H(z)} \nonumber\\
&&{=}\frac{c}{\sqrt{3}}\int_0^{1/(1+z)}\frac{da}{a^2H(a)\sqrt{1+(3\Omega_b/(4\Omega_\gamma)a)}},
\end{eqnarray}
where $c_s$ is the sound speed of the photon$-$baryon fluid
\cite{ref:Hu1, ref:Hu2, ref:Caldwell}:
\begin{eqnarray}
&&c_s^{-2}=3+\frac{4}{3}\times\frac{\rho_b(z)}{\rho_\gamma(z)}=3+\frac{4}{3}\times\left(\frac{\Omega_b}{\Omega_\gamma}\right)a,
\end{eqnarray}
and here $\Omega_\gamma=2.469\times10^{-5}h^{-2}$ for
$T_{CMB}=2.725K$.

\begin{table}[htbp]
\begin{center}
\begin{tabular}{c|l}
\hline\hline
 $z$ &\ $r_s(z_d)/D_V(z)$  \\ \hline
 $0.2$ &\ $0.1905\pm0.0061$  \\ \hline
 $0.35$  &\ $0.1097\pm0.0036$  \\
\hline
\end{tabular}
\end{center}
\caption{\label{baodata} The observational $r_s(z_d)/D_V(z)$
data~\cite{ref:Percival2}.}
\end{table}
Using the data of BAO in Table \ref{baodata} and the inverse
covariance matrix $V^{-1}$ in \cite{ref:Percival2}:

\begin{eqnarray}
&&V^{-1}= \left(
\begin{array}{cc}
 30124.1 & -17226.9 \\
 -17226.9 & 86976.6
\end{array}
\right),
\end{eqnarray}

thus, the $\chi^2_{BAO}(p_s)$ is given as
\begin{equation}
\chi^2_{BAO}(p_s)=X^tV^{-1}X,\label{eq:chi2BAO}
\end{equation}
where $X$ is a column vector formed from the values of theory minus
the corresponding observational data, with
\begin{eqnarray}
&&X= \left(
\begin{array}{c}
 \frac{r_s(z_d)}{D_V(0.2)}-0.190533 \\
 \frac{r_s(z_d)}{D_V(0.35)}-0.109715
\end{array}
\right),
\end{eqnarray}
and $X^t$ denotes its transpose.

\subsection{Cosmic Microwave Background constraints}

The CMB shift parameter $R$ is provided by \cite{ref:Bond1997}
\begin{equation}
R(z_{\ast})=\sqrt{\Omega_m H^2_0}(1+z_{\ast})D_A(z_{\ast})/c,
\end{equation}
which is related to the second distance ratio
$D_A(z_\ast)H(z_\ast)/c$ by a factor $\sqrt{1+z_{\ast}}$. The
redshift $z_{\ast}$ (the decoupling epoch of photons) is obtained
using the fitting function \cite{Hu:1995uz}
\begin{equation}
z_{\ast}=1048\left[1+0.00124(\Omega_bh^2)^{-0.738}\right]\left[1+g_1(\Omega_m
h^2)^{g_2}\right],
\end{equation}
where the functions $g_1$ and $g_2$ read
\begin{eqnarray}
g_1&=&0.0783(\Omega_bh^2)^{-0.238}\left(1+ 39.5(\Omega_bh^2)^{0.763}\right)^{-1},\\
g_2&=&0.560\left(1+ 21.1(\Omega_bh^2)^{1.81}\right)^{-1}.
\end{eqnarray}
In additional, the acoustic scale is related to the first distance
ratio, $D_A(z_\ast)/r_s(z_\ast)$, and is defined as
\begin{eqnarray}
&&l_A\equiv(1+z_{\ast})\frac{\pi D_A(z_{\ast})}{r_s(z_{\ast})}.
\end{eqnarray}

Using the data of $R, l_A, 100\Omega_bh^2$ and their covariance
matrix of $[R(z_\ast), l_A(z_\ast), 100\Omega_bh^2]$ referring to
\cite{ref:Komatsu2008,ref:Bueno Sanchez}, we can calculate the
likelihood $L$ as $\chi^2_{CMB}=-2\ln L$:
\begin{eqnarray}
&&\chi^2_{CMB}=\bigtriangleup d_i[Cov^{-1}(d_i,d_j)][\bigtriangleup
d_i]^t,
\end{eqnarray}
where $\bigtriangleup d_i=d_i-d_i^{data}$ is a row vector, and
$d_i=(R, l_A, 100\Omega_bh^2)$.

\subsection{Observational Hubble Data constraints}

The observational Hubble data are based on differential ages of the
galaxies \cite{ref:JL2002}. In \cite{ref:JVS2003}, Jimenez {\it et
al.} obtained an independent estimate for the Hubble parameter using
the method developed in \cite{ref:JL2002}, and used it to constrain
the EOS of dark energy. The Hubble parameter depending on the
differential ages as a function of redshift $z$ can be written in
the form of
\begin{equation}
H(z)=-\frac{1}{1+z}\frac{dz}{dt}.
\end{equation}
So, once $dz/dt$ is known, $H(z)$ is obtained directly
\cite{ref:SVJ2005}. By using the differential ages of
passively-evolving galaxies from the Gemini Deep Deep Survey (GDDS)
\cite{ref:GDDS} and archival data
\cite{ref:archive1,ref:archive2,ref:archive3,ref:archive4,ref:archive5,ref:archive6},
Simon {\it et al.} obtained $H(z)$ in the range of $0\lesssim z
\lesssim 1.8$ \cite{ref:SVJ2005}. The twelve observational Hubble
data from \cite{ref:0905,ref:0907} are list in Table
\ref{Hubbledata}.
\begin{table}[htbp]
\begin{center}
\begin{tabular}{c|llllllllllll}
\hline\hline
 $z$ &\ 0 & 0.1 & 0.17 & 0.27 & 0.4 & 0.48 & 0.88 & 0.9 & 1.30 & 1.43 & 1.53 & 1.75  \\ \hline
 $H(z)\ ({\rm km~s^{-1}\,Mpc^{-1})}$ &\ 74.2 & 69 & 83 & 77 & 95 & 97 & 90 & 117 & 168 & 177 & 140 & 202  \\ \hline
 $1 \sigma$ uncertainty &\ $\pm 3.6$ & $\pm 12$ & $\pm 8$ & $\pm 14$ & $\pm 17$ & $\pm 60$ & $\pm 40$
 & $\pm 23$ & $\pm 17$ & $\pm 18$ & $\pm 14$ & $\pm 40$ \\
\hline
\end{tabular}
\end{center}
\caption{\label{Hubbledata} The observational $H(z)$
data~\cite{ref:0905,ref:0907}.}
\end{table}
In addition, in \cite{ref:0807}, the authors took the BAO scale as a
standard ruler in the radial direction, obtaining three more
additional data: $H(z=0.24)=79.69\pm2.32, H(z=0.34)=83.8\pm2.96,$
and $H(z=0.43)=86.45\pm3.27$.

 The best fit values of the model parameters from
observational Hubble data \cite{ref:SVJ2005} are determined by
minimizing
\begin{equation}
\chi_{Hub}^2(p_s)=\sum_{i=1}^{15} \frac{[H_{th}(p_s;z_i)-H_{
obs}(z_i)]^2}{\sigma^2(z_i)},\label{eq:chi2H}
\end{equation}
where $p_s$ denotes the parameters contained in the model, $H_{th}$
is the predicted value for the Hubble parameter, $H_{obs}$ is the
observed value, $\sigma(z_i)$ is the standard deviation measurement
uncertainty, and the summation is over the $15$ observational Hubble
data points at redshifts $z_i$.

\end{document}